\documentclass{aa}
\usepackage{psfig}

\def\0614{4U\thinspace0614+091}

\begin{document}

\bigskip \bigskip

\thesaurus{06 (02.01.2;  08.09.2;
               08.14.1; 13.25.5)}

\title{BeppoSAX Observations of the Atoll X-Ray Binary
4U~0614+091}

\author{S.~Piraino,$^{(1,2,3)}$ A.~Santangelo,$^{(2)}$
E.C.~Ford,$^{(4)}$  P.~Kaaret$^{(1)}$}

\authorrunning{S.~Piraino et al.} 

\titlerunning{BeppoSAX Observations of 4U~0614+091} 

\offprints {Philip Kaaret, email: pkaaret@cfa.harvard.edu}

\institute{(1) {\it Harvard-Smithsonian Center for
Astrophysics, 60 Garden St.,  Cambridge, MA 02138, USA}\\ (2)
{\it IFCAI/CNR, Via Ugo La Malfa 153, 90146 Palermo, Italy}\\
(3) {\it Dipartimento di Scienze Fisiche ed Astronomiche, via
Archirafi 36, 90123 Palermo, Italy}\\ (4) {\it University of
Amsterdam, Kruislaan 403, 1098 SJ Amsterdam,  The
Netherlands}\\}

\date{Received; accepted}

\maketitle

\begin{abstract} 

We report the first simultaneous measurement of the broad
band X-ray (0.3-150~keV) spectrum of the neutron star x-ray
binary \0614.  Our data confirm the presence of a hard x-ray
tail that can be modeled as thermal Comptonization of
low-energy photons on electrons having a very high
temperature, greater than 220~keV, or as a non-thermal
powerlaw.  We detected a spectral feature that can be
interpreted as reprocessing, via Compton reflection, of the
direct emission by an optically-thick disk and found a
correlation between the photon index of the power-law tail
and the fraction of radiation reflected which is similar to
the correlation found for black hole candidate x-ray binaries
and Seyfert galaxies.

\end{abstract}


\section{Introduction}

The x-ray source 4U~0614+091 was discovered in the Uhuru
survey and was localized by Copernicus (\cite{willmore74}).
The source is an x-ray burster (\cite{swank78}) and, thus,
contains a neutron star.  The luminosity of the x-ray bursts
constrains the distance to be within 3~kpc (\cite{brandt92}).
Classification of \0614 as an atoll source was suggested by
Singh \& Apparao (\cite{singh94}) and confirmed by Mendez
et~al.  (\cite{mendez97}).  New study of the source has been
stimulated by the discovery of high frequency quasi-periodic
oscillations (QPOs) in the persistent emission of this source
(\cite{ford97}) and of a hard x-ray tail extending to at least
100~keV (\cite{ford96}).

Here, we report on observations of \0614 obtained with the
narrow field instruments (NFIs) on board the BeppoSAX
satellite (\cite{boella97}).  We have obtained the first
simultaneous measurement of the broad band x-ray spectrum of
\0614, extending from soft x-rays to hard x-rays (0.3--150
keV) and are able to place strong constraints on the spectrum
of the hard x-ray emission.  We find that the continuum hard
spectrum of \0614 is well described by a powerlaw or thermal
Comptonization with a reflection component.  We find that the
thermal Comptonization temperature and the lower bound on any
exponential cutoff of the powerlaw model are both greater
than 200~keV.

We also studied the relation between the reprocessed
component and the intrinsic component.  A correlation between
the strength of reflection and the intrinsic spectral slope
has been found for GX~339-4 (\cite{ueda94}) and recently for
Seyfert AGNs, 2 black holes candidate (BHC) X-ray binaries,
and 2 X-ray bursters (Zdziarski et~al.  \cite{zdziarski99}).
We find a similar relation for \0614.

\section{Observations and Analysis} 

We performed a joint BeppoSAX/RXTE observation of \0614 on
19-20 October 1998 for a total of 42.6~ks of on--source
observing time in BeppoSAX (\cite{piraino99}).  Here, we use
data from the four BeppoSAX Narrow Field Instruments (NFIs)
in overlapping energy bands selected to give good signal to
noise for this source:  the Low Energy Concentrator
Spectrometer (LECS) for 0.1--4~keV, the Medium Energy
Concentrator Spectrometer (MECS) for 1.8--10~keV, the High
Pressure Gas Scintillation Proportional Counter (HPGSPC) for
4--27~keV, and the Phoswich Detection System (PDS) for
12--200~keV.  LECS and MECS data were extracted in circular
regions centered on the source position using radii of 8' and
4' respectively, containing 95\% of the source flux.  The
spectra have been rebinned to have at least 30 counts per
channel, and the HPGSPC and PDS spectra were grouped using a
logarithmic grid. A normalization factor has been included to
account for the mismatch in the BeppoSAX instruments'
absolute flux calibration.  The fit values of relative
normalization are in good agreement with values typically
observed (\cite{cusumano98}).

For the total spectrum presented in Figures~\ref{resid} and
\ref{spectrum}, we summed all of the 42.6~ks of data obtained
in October 1998.  To study the correlation of spectral slope
with reflection, presented in Figure~\ref{corr}, we divided
this observations into two intervals, based on luminosity, and
also used data from an observation performed on 13-14 October
1996 available from the BeppoSAX archive, again divided into
two intervals.

\section{Spectral Results}

\begin{figure}
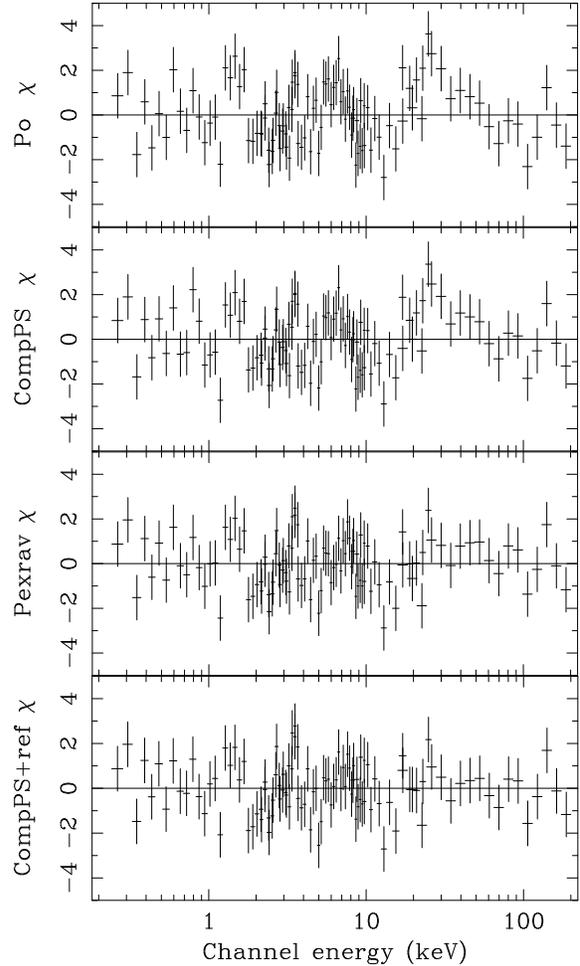

\centerline{\psfig{file=Bg023f1a.ps,height=3cm,rheight=2.95cm,angle=270.0}}
\centerline{\psfig{file=Bg023f1b.ps,height=3cm,rheight=2.95cm,angle=270.0}}
\centerline{\psfig{file=Bg023f1c.ps,height=3cm,rheight=2.95cm,angle=270.0}}
\centerline{\psfig{file=Bg023f1d.ps,height=3.85cm,rheight=3.85cm,angle=270.0}}
\caption{Residuals for the models described in Table~1.  For
display, the energy bins are equal to the $1\sigma$ energy
resolution.}
\label{resid} \end{figure}

\begin{figure}
\centerline{\psfig{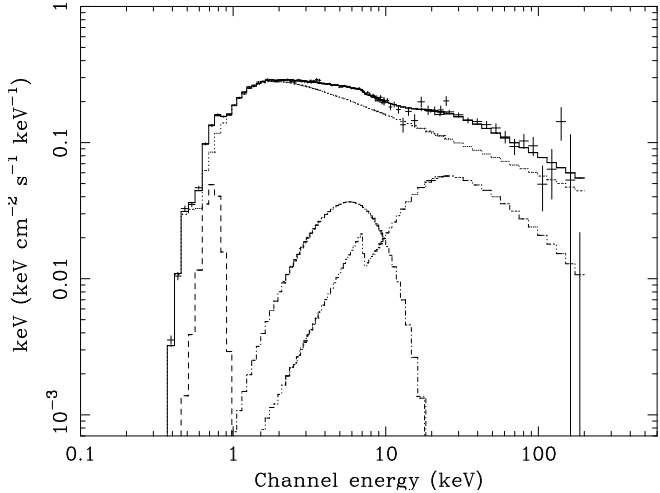}}
\caption{LECS, MECS, HPGSPC and PDS unfolded averaged
spectra, for the October 19--20, 1998 observation together
with a model consisting of a powerlaw, reflection, a
blackbody, and a low energy Gaussian line all with
absorption.  The total model fit is shown as a solid line,
the powerlaw as dotted line, the reflection component as a
dot-dot-dashed line, the blackbody component as a dot-dashed
line, and the Gaussian line as a dashed line.} 
\label{spectrum} \end{figure}

We fit the 0.1--200~keV BeppoSAX spectrum of \0614 with
several single component continuum models:  powerlaw,
powerlaw with an exponential high energy cutoff, three
different Comptonization models (\cite{sunyaev80};
\cite{titarchuk94}; \cite{Pout96a}), an exponentially cutoff
powerlaw and a Comptonization model (\cite{Pout96a}) each
with added reflection (\cite{magdziarz95}), thermal
bremsstrahlung, several disk models, and each of these single
component models with an added single temperature simple
blackbody component.  Each model also included absorption and
a Gaussian line near 0.7~keV (\cite{christian94};
\cite{white97}; \cite{schulz99}).  Only the powerlaw (with or
without an exponential cutoff) and Comptonization models gave
close to acceptable fits ($\chi^2_\nu\simeq$1.33) and in each
case the fit was improved by addition of the blackbody
component.  In Table~1 we report the best fit parameters for
the powerlaw model and the best fitting Comptonization model
(\cite{Pout96a}) both with and without reflection.  The fit
residuals of the models without reflection show a feature
between 10 and 60~keV that disappears when reflection is
added, see Figure~\ref{resid}.  The two models with
reflection provided the best fit with a significant
improvement in $\chi^2_\nu$ (see Table 1) relative to any
other model.

\begin{table}[t]  
\caption[]{Parameters for fits of the spectrum of \0614}
\label{tab:Tab2}  
\begin{flushleft}
\begin{tabular}{lllll} 
\hline \hline   \noalign{\smallskip} 
                      &\textbf{PL}    &\textbf{Comp} 
                      &\textbf{PL}    &\textbf{Comp}  \\  
                      &               &              
                      &\multicolumn{2}{l}{\textbf{with
                      reflection}} \\ 
\hline \hline 

$N_{\rm H}$     & 3.3$\pm$0.1   & 3.0$\pm$0.1   & 3.7$\pm$0.2   & 3.3$\pm$0.3 \\    
$E_{\rm g}$     & 0.71$\pm$0.03 & 0.70$\pm$0.03 & 0.66$\pm$0.04 & 0.65$\pm$0.05 \\
$\sigma$        & 0.07$\pm$0.02 & 0.07$\pm$0.02 & 0.09$\pm$0.02 & 0.11$\pm$0.02 \\
$EW$            & 88$\pm$33     & 90$\pm$20     & 140$\pm$64    & 140$\pm$60 \\ 
$\Gamma$        & 2.33$\pm$0.02 &               & 2.44$\pm$0.03 &  \\ 
$kT_{\rm s}$    &               & $< 0.04$      &               & $< 0.03$ \\
$kT_{\rm e}$    &               & 160$\pm$35    &               & 260$^{+50}_{-40}$ \\ 
$\tau$          &               & 0.26$\pm$0.1  &               & 0.10$\pm$0.03 \\  
$Refl$          &               &               & 0.90$\pm$0.25 & 0.45$\pm$0.25 \\
$kT_{\rm bb}$   & 1.45$\pm$0.08 & 1.28$\pm$0.15 & 1.47$\pm$0.06 & 1.39$\pm$0.06 \\ 
$F_{\rm bb}$    & 0.51$\pm$0.10 & 0.75$\pm$0.15 & 0.82$\pm$0.12 & 1.5$\pm$0.3 \\  
$F$             & 16.73         & 17.83         & 16.86         & 16.94  \\
$F_{\rm un}$    & 40.33         & 38.39         & 47.93         & 34.17  \\
$\chi^2_\nu (\nu$) & 1.21 (561) & 1.17(559)     & 1.09 (558)    & 1.08 (555) \\

\noalign{\smallskip} \hline

\end{tabular} \end{flushleft}  

{\small \sc Note} \small--- Given in the table are the
absorption column density ($N_{\rm H}$) [in units of
$10^{21}$ cm$^{-2}$], the centroid ($E_{\rm g}$) [keV], width
$(\sigma)$ [keV], and equivalent width $(EW)$ [keV] of the
low energy emission line, the photon index ($\Gamma$) of the
powerlaw, and the temperature of the input seed photon
distribution ($kT_{\rm s}$) [keV], the temperature ($kT_{\rm
e}$) [keV] and optical depth ($\tau$) of the Comptonizing
electron cloud, relative reflection fraction ($R$), the
temperature $(kT_{\rm bb})$ [keV] and flux $(F_{\rm bb})$
[$\rm 10^{-10} \, erg \, cm^{-2} \, s^{-1}$] of the
blackbody, the absorbed $(F)$ and unabsorbed $(F_{\rm un})$
flux [$\rm 10^{-10} \, erg \, cm^{-2} \, s^{-1}$ for
0.1--200~keV], and the reduced chi square, $\chi^2_\nu$, and
degrees of freedom ($\nu$).  All quoted errors represent
$90\%$ confidence level for a single parameter.

\end{table}

The powerlaw with reflection model is the PEXRAV/PEXRIV model
in XSPEC, which is an exponentially cut--off powerlaw
spectrum with reflection (\cite{magdziarz95}) from a disk. 
We assumed solar abundances and fixed at $60^{\circ}$ the
disk inclination angle.  Adding an exponential cutoff does
not improve the fit, either with or without reflection, and
the cutoff energy is well above 200~keV.  The Comptonization
model with reflection is COMPPS, kindly supplied by Juri
Poutanen.  This code calculates radiative transfer and
Comptonization in a two-phase disk-corona geometry
(\cite{Pout96a}).  Reflection is calculated via the same
method (\cite{magdziarz95}) as for the PEXRAV model.  We used
the model to calculate thermal Comptonization and used an
approximate treatment of radiative transfer using the escape
probability for a sphere for the results presented in
Table~1.  The temperature of the input soft photon
distribution is driven towards unexpectedly low values when
left as a free parameter.  We tried fixing this temperature
to that of the lowest temperature soft photon source present
in our spectral fit (the 0.27~keV blackbody), but were unable
to obtain good fits.  For both reflection models, the
ionization parameter is consistent with zero, but is not well
constrained.  For the results presented in Table~1 and
Figures~2 and 3, we used the PEXRAV model, which assumes a
neutral reflecting medium, with no exponential cutoff. 
Figure~2 shows the BeppoSAX data and the various components
of the fit using the PEXRAV model.

As the presence of an absorption edge is an inherent feature
of reflection, before accepting the reflection as component
of the model, we verified the presence in the spectrum of an
absorption edge.  Adding this feature to the powerlaw or
Comptonization models, we found an improvement in the
$\chi^2_\nu$ and the same edge energy and optical depth
($E_{\rm edge} = 8.5 \rm \, keV$, $\tau=0.1$) found by Singh
\& Apparao (\cite{singh94}) in a low state observation of
\0614 with EXOSAT.

Adding a Gaussian emission line to model iron K$_{\alpha}$
fluorescence did not improve the fit significantly and we
place upper limits of 20--60~eV on the equivalent width for
lines in the 6.4--8.1~keV range.  While iron K$_{\alpha}$
fluorescence is commonly associated with reflection in AGN,
the iron line emission found in Galactic BHC X-ray binaries
in the low (hard) state is much less than that found in AGN
(e.g. Zdziarski et al.\ \cite{zdziarski98}).  This may be due
to differences in the ionization state or geometry of the
disk, Doppler broadening and relativistic changes in the line
profile which make the line difficult to detect, or resonant
Auger destruction (\cite{Ross96}).  All of these effects may
also hold for neutron star X-ray binaries.  In addition, the
equivalent width of iron K$_{\alpha}$ for \0614 is likely to
be low since the equivalent width of iron K$_{\alpha}$
decreases approximately linearly with photon index $\Gamma$
as result of the decrease in the number of incident photons
with energy larger than the absorption edge energy for higher
$\Gamma$ (\cite{George91}; Vrtilek et al.~\cite{Vrtilek93})
and $\Gamma$ is larger for \0614 than for BHC x-ray binaries
in the low (hard) state.

We found a blackbody component with temperature, $kT_{\rm bb}$,
near 1.5~keV.  As the Comptonization optical depth is small,
the detection of the blackbody is consistent with an origin
within the Comptonization region.  We note that the blackbody
flux found using COMPPS is twice that found using the
powerlaw. Thus, it is unlikely that the blackbody component
is a spurious feature due to approximation of the
Comptonization spectrum with a powerlaw.  Blackbody
components have been found from \0614 in the past, with $kT$
ranging from 0.3~keV (\cite{christian94}; \cite{schulz99}) to
1.7~keV (Singh $\&$ Apparao \cite{singh94};
\cite{barret95}).  Adding a second blackbody component to our
model, we found that the second blackbody improves the fit
slightly and has $kT = 0.27 \pm 0.03 \rm \, keV$.  Thus, two
blackbody components may be simultaneously present in the
spectrum of this source. Addition of the second blackbody
does not significantly affect the other fit parameters.  We
also tried using the blackbody accretion disk models of
Shakura \& Sunayev (\cite{shakura73}), Stella \& Rosner
(\cite{stella84}), and Mitsuda (\cite{mitsuda84}).  These
models also gave good fits, but not better than the fits
obtained with the simple blackbody.

\begin{figure}
\centerline{\psfig{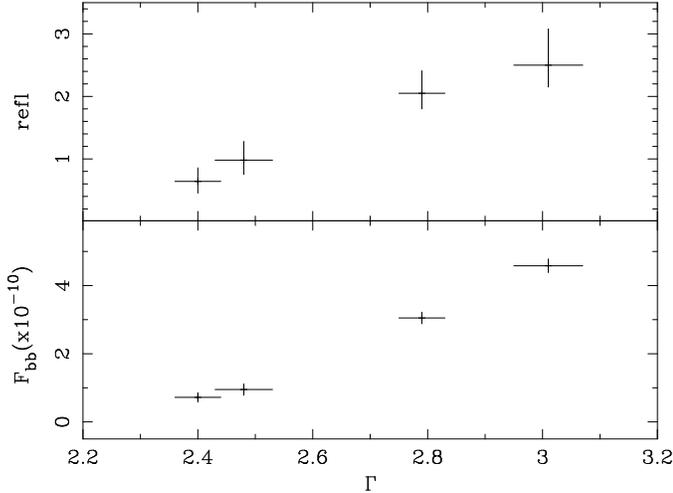}}
\caption{Spectral parameters correlations.  (a) Magnitude of
reflection versus photon index.  (b) Blackbody flux ($\rm
erg\; cm^{-2}\; s^{-1}$) versus photon index.}  \label{corr}
\end{figure}

Using the PEXRAV model in a total of four intervals over the
two observations, we found that the amount of reflection and
also the blackbody flux are correlated with the photon index
(Fig.~\ref{corr}).  The values found for the reflection
coefficient are large.  This may indicate that the photon
flux irradiating the disk is larger than the observed flux as
expected for relativistic electrons in an anisotropic
reflection geometry (\cite{ghisellini91}) or may simply
indicate that the inclination is less than $60^{\circ}$.  The
unabsorbed luminosity in the 0.1--200~keV band varied over
the range $5.1-29.1 \times 10^{36} (\frac {d}{3 \, \rm
kpc})^2 \, \rm erg \, s^{-1}$ in the four intervals.

\section{Discussion}

Before the BeppoSAX era, there were no x-ray burster
observations with simultaneous energy coverage from
0.1--200~keV.  This broad coverage is very important to
accurately measure the x-ray spectrum.  Using BeppoSAX, we
find a spectrum for the x-ray burster \0614 which can be
described either with a powerlaw with no detectable cutoff
below 200~keV or with a thermal Comptonization spectrum with
an electron temperature in excess of 220~keV.  As equally
good fits are obtained in the two cases, we cannot determine
whether the emission is due to a thermal or non-thermal
process.  If the emission is due to Comptonization, with
either a thermal or non-thermal electron energy distribution,
then the input soft photon distribution must peak at an
energy well below that of any observed soft photon source; in
particular, well below the temperature of the cooler
(0.27~keV) blackbody component.  Production of these soft
photons must be addressed in any successful application of a
Comptonization model to \0614.

The only robust evidence for the presence of a black hole in
an X-ray binary is a measurement of the mass of the compact
object indicating a value greater than $3 \rm \, M_{\odot}$.
While neutron stars have unambiguous x-ray observational
signatures, such as type I X-ray bursts or coherent X-ray
pulsations, black holes at best offer negative evidence: the
absence of any clear neutron star characteristics. Several
criteria have been proposed to distinguish black holes from
neutron stars in x-ray binaries based on solely on their
x-ray emission, but so far none has been found be truly
unique to black holes.  It has been proposed that
two-component spectra, with a thermal component and an
extended steep power-law component, are a signature of the
presence of a black hole in a binary system
(\cite{laurent99}).  Our data clearly show the presence of
both a soft thermal component and a steep power-law component
extending to at least 200~keV from an x-ray binary containing
a neutron star.  Again, a criterion proposed to distinguish
black hole versus neutron star binaries based on their x-ray
emission is found inadequate.  

The spectrum we observe from 4U~0614+091 may also be compared
to the proposal by Zdziarski et~al.\ (\cite{zdziarski98})
that very high energy cutoffs, $E_{\rm cut} \geq 100 \, \rm keV$,
and thermal Comptonization temperatures, $kT_{\rm e} \geq 50 \,
\rm keV$ are a signature for black holes.  However, this
criterion applies only to the hard state ($\Gamma < 2$)
emission, while  we found 4U~0614+091 in a softer state
($\Gamma = 2.45$). It would be of great interest to observe
4U~0614+091 with BeppoSAX while it is in a harder state to
see if the spectrum still extends to high energies without a
cutoff.

Our best deconvolution of the spectrum requires reflection
from an optically-thick disk.  Reflection models have worked
very well in fitting broad bumps, between 10--60~keV, visible
in the spectra of black hole candidates (\cite{done92};
\cite{Gierlinski97}; Zdziarski et al.\ \cite{zdziarski98}). 
For two other x--ray bursters, 4U~1608--522 and GS~1826--238,
observed by Ginga in their hard (low) state the reflection
model gave also good results (\cite{yoshida93}; \cite{
strickman96}).  The excellent broad band coverage of BeppoSAX
gives strong evidence for reflection in the spectrum of
\0614. We find that the magnitude of the reflection and the
flux of the blackbody component are both correlated with the
photon index.  Zdziarski et al.\ (\cite{zdziarski99}) found a
common correlation between strength of reflection and
spectral slope in Seyfert AGNs, 2 black hole candidate X-ray
binaries, and the X-ray bursters 4U~1608-52 and GS~1826-238. 
The correlation we find for 4U~0614+091 is similar in form
but may be slightly displaced to higher values of the photon
index. Additional high-quality broad-band spectra obtained
with BeppoSAX would permit more detailed study of this
correlation in the future.

\begin{acknowledgements}

We thank Juri Poutanen, Didier Barret, and Fred Lamb for
useful discussions, and the referee, Andrzej Zdziarski, for
comments which improved the paper.  This research has made use
of data obtained through the BeppoSAX archive, provided by
the BeppoSAX SDC team.  PK and SP acknowledge support from
NASA grants NAG5-7405 and NAG5-7334.

\end{acknowledgements}

\end{document}